\documentclass{article}

\usepackage{geometry}
\usepackage{comment} 
\usepackage{graphicx}
\usepackage{subfigure}
\usepackage{url}
\usepackage{amsfonts}
\usepackage{amsmath}
\usepackage{amsthm}
\usepackage{bm} 
\usepackage[usenames,dvipsnames]{color}
\usepackage{algorithm}
\usepackage{algorithmic}
\usepackage{tikz}
\usetikzlibrary{arrows}
\usepackage{arydshln}
\usepackage{mathtools}
\usepackage[numbers]{natbib}
\usepackage[colorlinks,pdftex,pdfpagelabels,bookmarks,hyperindex,hyperfigures]{hyperref}
\hypersetup{
  pdftitle={},
  pdfauthor={Reza Sameni},
    pdfnewwindow=true,      
    colorlinks=true,       
    linkcolor=blue,          
    citecolor=magenta,        
    filecolor=cyan,      
    urlcolor=teal
}
\DeclareMathOperator{\Tr}{tr}
\DeclareMathOperator{\Diag}{\mathbf{diag}}
\DeclareMathOperator*{\argmax}{arg\,max}
\DeclareMathOperator*{\argmin}{arg\,min}

\newtheorem{example}{Example}
\usepackage{mdframed}
\usepackage{lipsum}
\newmdtheoremenv{post}{Postulate}

\newtheorem{theorem}{Theorem}
\newtheorem{lemma}[theorem]{Lemma}

\begin{document}



\title{Notes on Information Propagation in Noisy Multichannel Data Models: Insights into Sensor Selection and Fusion in Multimodal Biomedical Applications}

\author{Reza~Sameni
\thanks{R.~Sameni is with the Department of Biomedical Informatics, Emory University (email: \url{rsameni@dbmi.emory.edu}).
}%
}
\date{December 2023}
\maketitle
\begin{abstract}

Multimodality and multichannel monitoring have become increasingly popular and accessible in engineering, Internet of Things, wearable devices, and biomedical applications. In these contexts, given the diverse and complex nature of data modalities, the relevance of sensor fusion and sensor selection is heightened. In this note, we study the problem of channel/modality selection and fusion from an information theoretical perspective, focusing on linear and nonlinear signal mixtures corrupted by additive Gaussian noise. We revisit and extend well-known properties of linear noisy data models in estimation and information theory, providing practical insights that assist in the decision-making process between channel (modality) selection and fusion. Using the notion of multichannel signal-to-noise ratio, we derive conditions under which, selection or fusion of multimodal/multichannel data can be beneficial or redundant. This contributes to a better understanding of how to optimize sensor fusion and selection from a theoretical standpoint, aiming to
enhance multimodal/multichannel system design, especially for biomedical multichannel/multimodal applications.
\end{abstract}

\section{Introduction}
\label{sec:introduction}
The problem of modality/channel selection has vast applications in biomedical research. This includes lead selection and fusion for noninvasive fetal \citep{Ghayem2021} and adult \citep{reyna2021will,reyna2022issues} electrocardiography; it also relates to the issue of multimodal cardiac monitoring \citep{kazemnejad2021open}.
Multimodal and multichannel data processing is motivated by the assumption that each modality conveys complementary information, leading to improved overall performance compared to single modality (or single-channel) approaches in terms of detection rate, estimation performance, classification rate, etc. However, despite expectations, some studies have shown that this assumption is not always accurate. When the data model is inaccurate, depending on the noise level and the correlations between the source signals and noises, selecting a particular modality (or channel) can sometimes be more effective than combining them. It has been shown that under imperfect data models, adding dimensions and modalities can actually degrade overall performance \citep{chlaily2017interaction,chlaily2018modele}. This phenomenon has been theoretically verified from an information and estimation theoretical perspective and is applicable even with as few as two data channels or modalities.

In this note, we aim to expand on these findings by further examining the estimation theoretical properties of multichannel/multimodal linear and nonlinear source-sensor mixtures corrupted by additive Gaussian noise. Using standard estimation schemes, including least squares, maximum likelihood and Bayesian estimation, we shown that the signal-to-noise-ratio (SNR) matrix \citep{Reeves2018} --- a generalization of the notion of scalar SNR --- plays a significant role in channel fusion and selection in all estimation schemes.

\section{Data model}
\label{sec:datamodel}
Consider the following data model:
\begin{equation}
    \bm{x} = \mathbf{A} \bm{s} + \bm{v} = \sum_{k = 1} ^n \bm{a}_k s_k + \bm{v}
\label{eq:linearmixture}
\end{equation}
where $\bm{x} = [x_1, \ldots, x_n]^T\in\mathbb{R}^{n}$ is the measurement/observation vector, $\bm{s} = [s_1, \ldots, s_m]^T\in\mathbb{R}^{m}$ represents the source vector with covariance matrix $\mathbf{P}=\mathbb{E}\{(\bm{s}-\mathbb{E}\{\bm{s}\})(\bm{s}-\mathbb{E}\{\bm{s}\})^T\} \in\mathbb{R}^{m\times m}$, $\bm{v} = [v_1, \ldots, v_n]\in\mathbb{R}^{n}$ is additive noise independent from $\bm{s}$, and $\mathbf{A} = [\bm{a}_1, \ldots, \bm{a}_m]\in\mathbb{R}^{n\times m}$ is the mixing matrix. 
An alternative row-wise (channel-wise) representation of the observations, which we will use is:
\begin{equation}
    x_k = \tilde{\bm{a}}_k^T \bm{s} + v_k
\label{eq:linearmixturechwise}
\end{equation}
where $\tilde{\bm{a}}_k^T$ is the $k$th row of the mixing matrix $\mathbf{A}$ ($k=1,\ldots,n$). 

\section{Source estimation}
Throughout the note, the objective is to estimate $\bm{s}$. We will investigate three levels of prior knowledge:
\begin{enumerate}
    \item Data model-only, using a \textit{least squares} formulation;
    \item Gaussian noise model presumption using a \textit{maximum likelihood} framework; and
    \item The fully \textit{Bayesian} framework, in which the data-model, noise distribution and the source distribution are all presumed.
\end{enumerate}
We will see that from top to bottom, while the actual measurements $\bm{x}$ are the same, we gain additional information through priors presumptions on the source and noise.

\subsection{Least squares}
\label{sec:LSestimate}
Considering the data model in \eqref{eq:linearmixture}, without any additional priors on the source or noise distributions, we have limited options to estimate the source vector. In this case, the \textit{least squares (LS)} and \textit{weighted least squares (WLS)} are the most popular:
\begin{equation}
\hat{\bm{s}}_{\text{WLS}} \stackrel{\Delta}{=} \argmin_{\bm{s}}{\|\bm{x} - \mathbf{A}\bm{s}\|_{\mathbf{W}}} = \argmin_{\bm{s}}{  (\bm{x} - \mathbf{A}\bm{s})^T\mathbf{W}(\bm{x} - \mathbf{A}\bm{s})}
\label{eq:WLSdef}
\end{equation}
where $\mathbf{W}\in\mathbb{R}^{n\times n}$ is a positive semi-definite matrix that to penalize the sample-wise measurement errors. The solution to the WLS problem is
\begin{equation}
\hat{\bm{s}}_{\text{WLS}} = (\mathbf{A}^T\mathbf{W}\mathbf{A})^{-1}\mathbf{A}^T\mathbf{W}\bm{x}
    \label{eq:WLSestimate}
\end{equation}
Setting $\mathbf{W}=\mathbf{I}$ simplifies to the standard least squares scheme, which handles all measurements with the same weight. Alternatively, if the measurement noise $\bm{v}$ is known to be zero-mean with covariance $\bm{\Sigma}$ (not necessarily Gaussian), we can intuitively set $\mathbf{W}=\bm{\Sigma}^{-1}$. This choice results in
\begin{equation}
\hat{\bm{s}}_{\text{WLS}} = (\mathbf{A}^T\bm{\Sigma}^{-1}\mathbf{A})^{-1}\mathbf{A}^T\bm{\Sigma}^{-1}\bm{x}
    \label{eq:LSestimate}
\end{equation}
Importantly, this is only a heuristic choice and the LS framework by itself does not justify this selection, beyond intuition, i.e., decorrelating the observation noises and down-weighting the noisier measurements. 

\subsubsection{Weighted least squares estimation quality}
\label{sec:WLSquality}
The WLS estimator can be evaluated in terms of its bias and error covariance. The WLS solution in \eqref{eq:LSestimate} is an \textit{unbiased estimate} of the deterministic or stochastic source vector $\bm{s}$. In other words, $\mathbb{E}\{\hat{\bm{s}}_{\text{WLS}}\}=\bm{s}$, leading to the error covariance matrix:
\begin{equation}
    \mathbf{C}_{\text{WLS}} \stackrel{\Delta}{=} \mathbb{E}\{(\bm{s} - \hat{\bm{s}}_{\text{WLS}})(\bm{s} - \hat{\bm{s}}_{\text{WLS}})^T\}= (\mathbf{A}^T\bm{\Sigma}^{-1}\mathbf{A})^{-1}
\label{eq:LSestimateCov}
\end{equation}
Accordingly, the estimation quality depends only on the matrix:
\begin{equation}
\textbf{\texttt{snr}}\stackrel{\Delta}{=}\mathbf{A}^T\bm{\Sigma}^{-1}\mathbf{A}
\label{eq:snr}
\end{equation}
This matrix is recognized in the literature as the \textit{signal to noise ratio (SNR) matrix} \citep{Reeves2018}, representing a multichannel extension of the conventional scalar SNR. To summarize, up to this point, we have only made two assumptions: i) the data model in \eqref{eq:linearmixture}, and ii) a zero-mean noise vector $\bm{v}$ with covariance matrix $\bm{\Sigma}$ and an arbitrary distribution.

\subsection{Maximum likelihood}
\label{sec:ML}
The LS formulation did not require (and benefit from) any assumptions on the distributions of the source and noise vectors. If any such priors are available, one may use the \textit{maximum likelihood (ML)} framework, which seeks: 
\begin{equation}
    \hat{\bm{s}}_{\text{ML}}(\bm{x}) \stackrel{\Delta}{=}\argmax_{\bm{s}}{
    f_{\bm{X}|\bm{S}}(\bm{x}|\bm{s})}
    \label{eq:MLestim1}
\end{equation}
which reads: \textit{``find the source vector that maximizes the probability of observing a given $\bm{x}$.''} To answer this question, the noise distribution is required, while the source $\bm{s}$ can be deterministic or have an arbitrary stochastic distribution. For example, if the noise is Gaussian $\bm{v}\sim\mathcal{N}(\mathbf{0}, \bm{\Sigma})$, we find:
\begin{equation}
\begin{array}{ll}
    \hat{\bm{s}}_{\text{ML}}(\bm{x}) &=\displaystyle\argmax_{\bm{s}}{
    f_{\bm{X}|\bm{S}}(\bm{x}|\bm{s})}=\argmax_{\bm{s}}{
    f_{\bm{v}}(\bm{x}-\mathbf{A}\bm{s})} \\&= \displaystyle\argmax_{\bm{s}}{  (2\pi)^{-n/2}\det(\bm{\Sigma})^{-1/2}\exp[-\frac{1}{2}(\bm{x}-\mathbf{A}\bm{s})^T\bm{\Sigma}^{-1}(\bm{x}-\mathbf{A}\bm{s})]}\\&= (\mathbf{A}^T\bm{\Sigma}^{-1}\mathbf{A})^{-1}\mathbf{A}^T\bm{\Sigma}^{-1}\bm{x}
\end{array}    
    \label{eq:MLestim2}
\end{equation}
which is the same solution obtained from the intuitive WLS formulation.

\subsubsection{Maximum likelihood estimation quality}
\label{sec:MLquality}
Inserting the ML estimator derived in \eqref{eq:MLestim2} into the data model \eqref{eq:linearmixture}, we find:
\begin{equation}
    \hat{\bm{s}}_{\text{ML}}(\bm{x}) = (\mathbf{A}^T\bm{\Sigma}^{-1}\mathbf{A})^{-1}\mathbf{A}^T\bm{\Sigma}^{-1}\bm{x} = \bm{s} + (\mathbf{A}^T\bm{\Sigma}^{-1}\mathbf{A})^{-1}\mathbf{A}^T\bm{\Sigma}^{-1}\bm{v}
    \label{eq:MLestimateexpanded}
\end{equation}
Therefore $\mathbb{E}\{\hat{\bm{s}}_{\text{ML}}(\bm{x})\}=\bm{s}$, i.e., the estimator is \textit{unbiased}, and the error covariance of the ML estimator is:
\begin{equation}
    \mathbf{C}_{\text{ML}} \stackrel{\Delta}{=} \mathbb{E}\{(\bm{s} - \hat{\bm{s}}_{\text{ML}})(\bm{s} - \hat{\bm{s}}_{\text{ML}})^T\}= (\mathbf{A}^T\bm{\Sigma}^{-1}\mathbf{A})^{-1}=\textbf{\texttt{snr}}^{-1}
    \label{eq:MLestimateCov}
\end{equation}
which is the inverse of the SNR matrix defined in \eqref{eq:snr}.

\subsection{Bayesian estimation framework}
Bayesian estimation is the most complete of the three frameworks, which benefits from the data model, noise and source distributions. Bayesian estimation generally seeks
\begin{equation}
\hat{\bm{s}}_{\text{bys}} = \argmin_{\hat{\bm{s}}} \mathbb{E}\{c(\bm{s}, \hat{\bm{s}})\}
\label{eq:bayesian}
\end{equation}
where the expectation is taken over the joint distribution $f_{\bm{X},\bm{S}}(\bm{x},\bm{s})$, and $c(\bm{s}, \hat{\bm{s}})$ is a non-negative cost of error typically chosen to be in the form of $c(\bm{s}, \hat{\bm{s}})=L(|\bm{s}- \hat{\bm{s}}|)$ fulfilling $L(0) = 0$ \citep{VanTrees2001detection}. 

The \textit{minimum mean square error (MMSE)} estimator is a special case for the Bayesian estimator \eqref{eq:bayesian}, which uses a quadratic error function  
$c(\bm{s}, \hat{\bm{s}}) \stackrel{\Delta}{=}(\bm{s}- \hat{\bm{s}})^T\mathbf{M}(\bm{s}- \hat{\bm{s}})$, 
where $\mathbf{M}$ is an arbitrary positive semi-definite matrix. This choice leads to the MMSE estimator:
\begin{equation}
    \hat{\bm{s}}_{\text{mmse}} \stackrel{\Delta}{=}
    \mathbb{E}\{\bm{s}|\bm{x}\}
    \label{eq:MMSEestimate}
\end{equation}
also known as the conditional mean. More generally, under a set of sufficient conditions --- originally derived by S.~Sherman --- $\hat{\bm{s}}_{\text{mmse}}$ is also the optimal estimator for any other symmetric convex error cost function $c(\bm{s}, \hat{\bm{s}})$ \citep[Ch. 2.4]{VanTrees2001detection}.


\subsubsection{The MMSE estimation quality}
\label{sec:MMSEquality}
As noted above, the MMSE estimator is \textit{unbiased} and the error covariance matrix is:
\begin{equation}
    \mathbf{C}_{\text{mmse}} \stackrel{\Delta}{=} \mathbb{E}\{(\bm{s} - \hat{\bm{s}}_{\text{mmse}})(\bm{s} - \hat{\bm{s}}_{\text{mmse}})^T\}
    \label{eq:MMSEestimateCov}
\end{equation}

\subsection{The Cram\'er Rao lower bound}
\label{sec:MLCRLB}
It is insightful to review the \textit{Cram\'er Rao lower bound (CRLB)} for the data model \eqref{eq:linearmixture}, in the Bayesian and maximum likelihood frameworks.
\subsubsection{The CRLB for deterministic (fixed) sources}
Supposing that $\hat{\bm{s}}(\bm{x})$ is an unbiased estimator of the unknown vector $\bm{s}$ (deterministic or stochastic but fixed), there exists a minimum on the covariance of the estimation error \citep[pp. 79--80]{VanTrees2001detection}. Accordingly\footnote{Note that the matrix inequality $\mathbf{A}\succeq \mathbf{B}$, used in \eqref{eq:CRLBml}, denotes that $\mathbf{A}-\mathbf{B}$ is \textit{positive semi-definite}.},
\begin{equation}
\mathbb{E}_{\bm{x}}\{[\bm{s} - \hat{\bm{s}}(\bm{x})][\bm{s} - \hat{\bm{s}}(\bm{x})]^T\} \succeq \mathbf{J}_{\bm{x}|\bm{s}}^{-1}
    \label{eq:CRLBml}
\end{equation}
where the expectation is over the observations $\bm{x}$ and $\mathbf{J}_{\bm{x}|\bm{s}}$ (which is generally a function of $\bm{s}$) is the \textit{Fisher information matrix} with the following entries:
\begin{equation}
[\mathbf{J}_{\bm{x}|\bm{s}}]_{kl} \stackrel{\Delta}{=}\mathbb{E}_{\bm{x}}\{\frac{\partial}{\partial s_k}\log f_{\bm{X}|\bm{S}}(\bm{x}|\bm{s}) \frac{\partial}{\partial s_l}\log f_{\bm{X}|\bm{S}}(\bm{x}|\bm{s})\} = - \mathbb{E}_{\bm{x}}\{\frac{\partial^2}{\partial s_k s_l}\log f_{\bm{X}|\bm{S}}(\bm{x}|\bm{s})\}
    \label{eq:fisherinformation}
\end{equation}
An unbiased estimator that reaches the CRLB is called an \textit{efficient estimator}, and whenever an efficient estimator exists, it is the ML estimator.

We can show that for the data model defined in \eqref{eq:linearmixture} and a Gaussian noise:
\begin{equation}
\mathbf{J}_{\bm{x}|\bm{s}}=\mathbf{A}^T\bm{\Sigma}^{-1}\mathbf{A}=\textbf{\texttt{snr}}
\label{eq:fisherinformationGaussian}
\end{equation}
In other words, for the data model of interest, the SNR matrix is equal to the Fisher information matrix and the ML estimate is an efficient estimator (compare \eqref{eq:MLestimateCov} with \eqref{eq:fisherinformationGaussian}).

It should be noted that the ML estimator does not necessarily exist, for instance in the cases that there is no information in the observations regarding an entry (or some entries) of the source vector. In the linear model, a necessary condition for the existence of the ML estimate is that the observation vector length be equal to or greater than the source vector length (otherwise the Fisher information matrix is singular).

\subsubsection{The CRLB for stochastic sources}
For random source vectors, a bound similar to the CRLB exists for the lower bound of the covariance error. According to \citep[p. 72, 84]{VanTrees2001detection}, the definition of the CRLB is similar to \eqref{eq:CRLBml}:
\begin{equation}
\mathbb{E}\{[\bm{s} - \hat{\bm{s}}(\bm{x})][\bm{s} - \hat{\bm{s}}(\bm{x})]^T\} \succeq \mathbf{J}_{\bm{x}}^{-1}
    \label{eq:CRLBrandomvar}
\end{equation}
where the expectations are calculated over the joint distribution $f_{\bm{X},\bm{S}}(\bm{x},\bm{s})$, and the information matrix has the following entries:
\begin{equation}
[{\mathbf{J}_{\bm{x}}}]_{kl} \stackrel{\Delta}{=}\mathbb{E}\{\frac{\partial}{\partial s_k}\log f_{\bm{X},\bm{S}}(\bm{x},\bm{s}) \frac{\partial}{\partial s_l}\log f_{\bm{X},\bm{S}}(\bm{x},\bm{s})\} = - \mathbb{E}\{\frac{\partial^2}{\partial s_k s_l}\log f_{\bm{X},\bm{S}}(\bm{x},\bm{s})\}
\label{eq:fisherinformationRandom}
\end{equation}
Using Bayes' rule, the information matrix of this random case can be related to the Fisher information as follows \citep[p. 84]{VanTrees2001detection}:
\begin{equation}
{\mathbf{J}_{\bm{x}}} = \mathbf{J}_{\bm{x}|\bm{s}} + \mathbf{J}_{\bm{s}}
\label{eq:fisherinformationRandom2}
\end{equation}
i.e., the \textit{total information} matrix is equal to the information obtained by the measurements (subject to a presumed source vector) plus the \textit{prior information} $\mathbf{J}_{\bm{s}}$, which is entry-wise defined as follows:
\begin{equation}
[\mathbf{J}_{\bm{s}}]_{kl} \stackrel{\Delta}{=}\mathbb{E}_{\bm{s}}\{\frac{\partial}{\partial s_k}\log f_{\bm{S}}(\bm{s}) \frac{\partial}{\partial s_l}\log f_{\bm{S}}(\bm{s})\} = - \mathbb{E}_{\bm{s}}\{\frac{\partial^2}{\partial s_k s_l}\log f_{\bm{S}}(\bm{s})\}
\label{eq:priorinformation}
\end{equation}
and is only a function of the source distribution.

Therefore, for the data model \eqref{eq:linearmixture}, we conclude the following lemma 
\begin{center}
\noindent\fbox{%
    \parbox{\textwidth}{%
\begin{lemma}
\label{theorem:CRLB}
For the data model $\bm{x}=\mathbf{A}\bm{s}+\bm{v}$, with $\bm{v}\sim\mathcal{N}(\mathbf{0}, \bm{\Sigma})$ independent from $\bm{s}$, the CRLB for an arbitrary unbiased estimator $\hat{\bm{s}}(\bm{x})$ is
\begin{equation}
\mathbb{E}\{[\bm{s} - \hat{\bm{s}}(\bm{x})][\bm{s} - \hat{\bm{s}}(\bm{x})]^T\} \succeq \mathbf{J}_{\bm{x}}^{-1}
    \label{eq:CRLBGaussianChannel}
\end{equation}
where
\begin{equation}
    {\mathbf{J}_{\bm{x}}} = 
    \textbf{\texttt{snr}} + \mathbf{J}_{\bm{s}}
\label{eq:totalinformationMatrix}
\end{equation}
which shows the additive property between the Fisher information (represented by the mixture SNR matrix  $\textbf{\texttt{snr}}=\mathbf{A}^T\bm{\Sigma}^{-1}\mathbf{A}$) and the prior information regarding the source vector, defined as:
\begin{equation}
\mathbf{J}_{\bm{s}} \stackrel{\Delta}{=}- \mathbb{E}_{\bm{s}}\{\nabla_{\bm{s}}(\nabla_{\bm{s}}^T\log f_{\bm{S}}(\bm{s}))\}
\label{eq:priorinformationMatrix}
\end{equation}
\end{lemma}
}%
}
\end{center}

\begin{example}[Gaussian distributed sources]
    As a special case, for a source vector $\bm{s}$ with a Gaussian distribution $\bm{s}\sim\mathcal{N}(\bm{\mu}, \bm{\Gamma})$, we obtain $\mathbf{J}_{\bm{s}} = \bm{\Gamma}^{-1}$, resulting in: ${\mathbf{J}_{\bm{x}}} = \textbf{\texttt{snr}} + \bm{\Gamma}^{-1}$.
\end{example}


\subsection{Fisher information for two set of sensors}
\label{sec:crlbtwogroup}
Let us now expand the previous results to two sets of sensors from the same vector source:
\begin{equation}
\begin{array}{cc}
     \bm{x} = \mathbf{A} \bm{s} + \bm{v}\\
     \bm{y} = \mathbf{B} \bm{s} + \bm{u}
\end{array}
\label{eq:twolinearmixturesRepeated}
\end{equation}
where $\bm{v}\sim\mathcal{N}(\mathbf{0}, \bm{\Sigma}_v)$ and $\bm{u}\sim\mathcal{N}(\mathbf{0}, \bm{\Sigma}_u)$ are jointly Gaussian. Therefore, the augmented observation vector $\bm{z}\stackrel{\Delta}{=}[\bm{x}^T, \bm{y}^T]^T$ is also Gaussian distributed: $\bm{z}\sim\mathcal{N}(\mathbf{0}, \bm{\Sigma})$, with
\begin{equation}
\bm{\Sigma} \stackrel{\Delta}{=} \left[
\begin{array}{cc}
     \bm{\Sigma}_v & \bm{\Sigma}_{vu}\\
     \bm{\Sigma}_{uv} & \bm{\Sigma}_u
\end{array}\right]
\label{eq:noisecovariance}
\end{equation}
Using Lemma \ref{theorem:CRLB}, we can obtain the information matrices for each and the augmented observations:
\begin{equation}
\begin{array}{rl}
    \mathbf{J}_{\bm{x}} = & \mathbf{A}^T\bm{\Sigma}_v^{-1}\mathbf{A} + \mathbf{J}_{\bm{s}}\\
    \mathbf{J}_{\bm{y}} = & \mathbf{B}^T\bm{\Sigma}_u^{-1}\mathbf{B} + \mathbf{J}_{\bm{s}}\\
    \mathbf{J}_{\bm{x},\bm{y}} = & \mathbf{A}^T\bm{\Sigma}_{11}^{-1}\mathbf{A} + \mathbf{A}^T\bm{\Sigma}_{12}^{-1}\mathbf{B} + \mathbf{B}^T\bm{\Sigma}_{21}^{-1}\mathbf{A} + \mathbf{B}^T\bm{\Sigma}_{22}^{-1}\mathbf{B} + \mathbf{J}_{\bm{s}} 
\end{array}
\label{eq:InfoFunTwoLinMix}
\end{equation}
where $\bm{\Sigma}_{ij}^{-1}$ ($i,j \in \{1, 2\}$) denote the four block entries of $\bm{\Sigma}^{-1}$ with appropriate dimensions. Considering that $\mathbf{J}_{\bm{s}}$ is a common term in all cases, we conclude that the source distribution does not influence the information difference between two sets of measurements, as shown in the sequel.

\begin{example}[CRLB of two-channel recordings]
Consider two scalar recordings $x_i = \bm{a}_i^T \bm{s} + v_i$ $(i = 1, 2)$, with $v_i\sim\mathcal{N}(0, \sigma_i^2)$. From (\ref{eq:InfoFunTwoLinMix}) we find:
\begin{equation}
    \mathbf{J}_{x_i} = \displaystyle\frac{\bm{a}_i\bm{a}_i^T}{\sigma_i^2} + \mathbf{J}_{\bm{s}}    
\label{eq:twosensorsInfo}
\end{equation}
and for the concatenated set of observations (generally with correlated observation noises), defining $(v_1, v_2)^T\sim\mathcal{N}(0, \bm{\Sigma})$, where $\bm{\Sigma} \stackrel{\Delta}{=} \left(\begin{array}{cc}
     \sigma_{11} & \sigma_{12} \\
     \sigma_{12} & \sigma_{22}
\end{array}\right)$ and $\sigma_{ii} \stackrel{\Delta}{=} \sigma_i^2$, we find:
\begin{equation}
    \mathbf{J}_{x_1,x_2} = \frac{1}{\sigma_{11}\sigma_{22} - \sigma_{12}^2}
    \left(\sigma_{22}\bm{a}_1\bm{a}_1^T - \sigma_{12}\bm{a}_2\bm{a}_1^T - \sigma_{12}\bm{a}_1\bm{a}_2^T + \sigma_{11}\bm{a}_2\bm{a}_2^T\right) + \mathbf{J}_{\bm{s}} 
\label{eq:twosensorsInfo2}
\end{equation}
If the noises are uncorrelated ($\sigma_{12}=0$), (\ref{eq:twosensorsInfo2}) simplifies to:
\begin{equation}
    \mathbf{J}_{x_1,x_2} = \displaystyle\frac{\bm{a}_1\bm{a}_1^T}{\sigma_1^2} + \displaystyle\frac{\bm{a}_2\bm{a}_2^T}{\sigma_2^2} + \mathbf{J}_{\bm{s}} 
\label{eq:twosensorsInfo2Simplified}
\end{equation}
Therefore, considering that $\bm{a}_i\bm{a}_i^T/\sigma_i^2$ are rank-one semi-positive definite matrices (with $m-1$ zero eigenvalues and a single non-zero eigenvalue equal to $\|\bm{a}_i\|/\sigma_i^2$), it is evident that $\mathbf{J}_{x_1,x_2} \succeq \mathbf{J}_{x_1}$ and $\mathbf{J}_{x_1,x_2} \succeq \mathbf{J}_{x_2}$, implying that using both observations is more informative than using any of the two. 
\end{example}

\subsubsection{Fisher information of augmented sensor sets}
We now seek a more compact representation form the third identity in \eqref{eq:InfoFunTwoLinMix}, i.e., the Fisher information of augmented sensor set $\bm{z}$. According to the block matrix inversion lemma, for arbitrary matrices $\mathbf{a}$, $\mathbf{b}$, $\mathbf{c}$ and $\mathbf{d}$ with appropriate dimensions ($\mathbf{a}$ and $\mathbf{d}$ should be square matrices) we have
\begin{equation}
\begin{bmatrix}
    \mathbf{a} & \mathbf{b} \\
    \mathbf{c} & \mathbf{d}
  \end{bmatrix}^{-1} = \begin{bmatrix}
     \mathbf{a}^{-1} + \mathbf{a}^{-1}\mathbf{b}\left(\mathbf{d} - \mathbf{ca}^{-1}\mathbf{b}\right)^{-1}\mathbf{ca}^{-1} &
      -\mathbf{a}^{-1}\mathbf{b}\left(\mathbf{d} - \mathbf{ca}^{-1}\mathbf{b}\right)^{-1} \\
    -\left(\mathbf{d} - \mathbf{ca}^{-1}\mathbf{b}\right)^{-1}\mathbf{ca}^{-1} &
       \left(\mathbf{d} - \mathbf{ca}^{-1}\mathbf{b}\right)^{-1}
  \end{bmatrix}
\label{eq:blockmatrixinversionlemma1}
\end{equation}
or alternatively
\begin{equation}
\begin{bmatrix}
    \mathbf{a} & \mathbf{b} \\
    \mathbf{c} & \mathbf{d}
  \end{bmatrix}^{-1} = \begin{bmatrix}
     \left(\mathbf{a} - \mathbf{bd}^{-1}\mathbf{c}\right)^{-1} &
      -\left(\mathbf{a}-\mathbf{bd}^{-1}\mathbf{c}\right)^{-1}\mathbf{bd}^{-1} \\
    -\mathbf{d}^{-1}\mathbf{c}\left(\mathbf{a} - \mathbf{bd}^{-1}\mathbf{c}\right)^{-1} &
       \quad \mathbf{d}^{-1} + \mathbf{d}^{-1}\mathbf{c}\left(\mathbf{a} - \mathbf{bd}^{-1}\mathbf{c}\right)^{-1}\mathbf{bd}^{-1}
  \end{bmatrix}
\label{eq:blockmatrixinversionlemma2}
\end{equation}
Using these identities in \eqref{eq:noisecovariance} and the symmetry of $\bm{\Sigma}$ ($\bm{\Sigma}_{uv} = \bm{\Sigma}_{vu}^T$), the block elements of $\bm{\Sigma}^{-1}$ can be obtained. It is therefore straightforward to calculate $\mathbf{J}_{\bm{x},\bm{y}}$ from the third identity in \eqref{eq:InfoFunTwoLinMix} in the following identical forms:
\begin{equation}
\begin{array}{l}
    \mathbf{J}_{\bm{x},\bm{y}} = \mathbf{A}^T\bm{\Sigma}_{v}^{-1}\mathbf{A} + (\mathbf{A}^T\bm{\Sigma}_{v}^{-1}\bm{\Sigma}_{vu} - \mathbf{B}^T)\mathbf{F}(\mathbf{A}^T\bm{\Sigma}_{v}^{-1}\bm{\Sigma}_{vu} - \mathbf{B}^T)^T + \mathbf{J}_{\bm{s}}\\
    \mathbf{J}_{\bm{x},\bm{y}} = \mathbf{B}^T\bm{\Sigma}_{u}^{-1}\mathbf{B} + (\mathbf{B}^T\bm{\Sigma}_{u}^{-1}\bm{\Sigma}_{uv} - \mathbf{A}^T)\mathbf{G}(\mathbf{B}^T\bm{\Sigma}_{u}^{-1}\bm{\Sigma}_{uv} - \mathbf{A}^T)^T + \mathbf{J}_{\bm{s}}
\end{array}
\label{eq:InfoFunTwoLinMixSimplified}
\end{equation}
where 
\begin{equation}
\begin{array}{l}
    \mathbf{F} \stackrel{\Delta}{=} (\bm{\Sigma}_{u} - \bm{\Sigma}_{uv}\bm{\Sigma}_{v}^{-1}\bm{\Sigma}_{vu})^{-1}\\
    \mathbf{G} \stackrel{\Delta}{=} (\bm{\Sigma}_{v} - \bm{\Sigma}_{vu}\bm{\Sigma}_{u}^{-1}\bm{\Sigma}_{uv})^{-1}    
\end{array}
\label{eq:FGMatrices}
\end{equation}
are the inverses of the so-called \textit{Schur components} of the block matrix $\bm{\Sigma}$. For covariance matrices --- which are positive definite --- it has been shown that the Schur components are also positive definite \citep[Appendix 5.5]{boyd2004convex}, \citep{gallier2010schur}.

Equations \eqref{eq:InfoFunTwoLinMix} and \eqref{eq:InfoFunTwoLinMixSimplified} can be combined to obtain:
\begin{equation}
\begin{array}{l}
    \mathbf{S}_{\bm{x},\bm{y}}^{\bm{x}} \stackrel{\Delta}{=}\mathbf{J}_{\bm{x},\bm{y}} - \mathbf{J}_{\bm{x}}=   (\mathbf{A}^T\bm{\Sigma}_{v}^{-1}\bm{\Sigma}_{vu} - \mathbf{B}^T)\mathbf{F}(\mathbf{A}^T\bm{\Sigma}_{v}^{-1}\bm{\Sigma}_{vu} - \mathbf{B}^T)^T\\
    \mathbf{S}_{\bm{x},\bm{y}}^{\bm{y}} \stackrel{\Delta}{=}\mathbf{J}_{\bm{x},\bm{y}} - \mathbf{J}_{\bm{y}}=   (\mathbf{B}^T\bm{\Sigma}_{u}^{-1}\bm{\Sigma}_{uv} - \mathbf{A}^T)\mathbf{G}(\mathbf{B}^T\bm{\Sigma}_{u}^{-1}\bm{\Sigma}_{uv} - \mathbf{A}^T)^T
\end{array}
\label{eq:InfoFunTwoVsOne}
\end{equation}
which are quadratic forms of the matrices $\mathbf{F}$ and $\mathbf{G}$. On the other hand, according to the \textit{Sylvester's law of inertia}, in real quadratic forms such as \eqref{eq:InfoFunTwoVsOne}, the left hand sides have the same number of positive, negative and zero eigenvalues as $\mathbf{F}$ and $\mathbf{G}$, respectively (if the multiplicand matrices are nonsingular). Therefore, we find $\mathbf{S}_{\bm{x},\bm{y}}^{\bm{x}}\succ \mathbf{0}$ and $\mathbf{S}_{\bm{x},\bm{y}}^{\bm{y}} \succ \mathbf{0}$.

Herein, we name these matrices \textit{synergic information matrices}, as they convey the excess information carried by the joint observations, as compared to using any of the observation sets individually. Apparently, since $\mathbf{S}_{\bm{x},\bm{y}}^{\bm{x}}$ and $\mathbf{S}_{\bm{x},\bm{y}}^{\bm{y}}$ are positive definite matrices, the Fisher information matrix of the joint measurements is always greater than or equal to the individual Fisher information matrices, and the CRLB is equal or lower when we use both modalities (observation sets), as compared to any of the two modalities. The boundary case is when the multiplicand factors in \eqref{eq:InfoFunTwoVsOne} are zero, which makes one of the modalities redundant. This case is only achievable with correlated noises $\bm{v}$ and $\bm{u}$.

\subsubsection{Prewhitening}
\label{sec:prewhitening}
A more simplified representation of the synergy between the sensor sets can be obtained by applying prewhitening on the two set of observations. Decomposing the measurement noise covariance matrices as $\bm{\Sigma}_{u}=\bm{L}_u\bm{L}_u^T$ and $\bm{\Sigma}_{v}=\bm{L}_v\bm{L}_v^T$, where $\bm{L}_u$ and $\bm{L}_v$ are also symmetric semi positive definite\footnote{A symetric semi positive definite matrix $\bm{A}$ can be uniquely decomposed as $\bm{A}=\bm{S}\bm{S}^T$, where $\bm{S}$ is symmetric semi positive definite, known as the square roots of $\bm{A}$ \cite[Ch.~6]{Strang2005}.}, the whitened version of the observations \eqref{eq:twolinearmixturesRepeated}
 are defined as:
\begin{equation}
\begin{array}{l}
\tilde{\bm{x}}=\bm{L}_v^{-1}\bm{x}=\bm{L}_v^{-1}\mathbf{A}\bm{s}+\bm{L}_v^{-1}\bm{v} = \tilde{\mathbf{A}}\bm{s}+\tilde{\bm{v}}\\
\tilde{\bm{y}}=\bm{L}_u^{-1}\bm{y}=\bm{L}_u^{-1}\mathbf{B}\bm{s}+\bm{L}_u^{-1}\bm{u} = \tilde{\mathbf{B}}\bm{s}+\tilde{\bm{u}}
\end{array}
\label{eq:whitened}
\end{equation}
where $\tilde{\bm{v}}\sim\mathcal{N}(\mathbf{0}, \mathbf{I})$ and $\tilde{\bm{u}}\sim\mathcal{N}(\mathbf{0}, \mathbf{I})$ are jointly Gaussian with the cross correlation matrix $\bm{\rho}\stackrel{\Delta}{=}\mathbb{E}(\tilde{\bm{v}}\tilde{\bm{u}})=\bm{L}_v^{-1}\bm{\Sigma}_{vu}\bm{L}_u^{-T}$ (or equivalently $\bm{\rho}^T=\mathbb{E}(\tilde{\bm{u}}\tilde{\bm{v}})=\bm{L}_u^{-1}\bm{\Sigma}_{uv}\bm{L}_v^{-T}$). With these notations, the information matrices defined in \eqref{eq:InfoFunTwoLinMixSimplified} can be rewritten as follows (for the original unwhitened observations):
\begin{equation}
\begin{array}{ll}
    \mathbf{J}_{\bm{x},\bm{y}} & =  (\tilde{\mathbf{A}}^T\bm{\rho} - \tilde{\mathbf{B}}^T)(\mathbf{I} -\bm{\rho}^T\bm{\rho} )^{-1}(\tilde{\mathbf{A}}^T\bm{\rho} - \tilde{\mathbf{B}}^T)^T + \tilde{\mathbf{A}}^T\tilde{\mathbf{A}} + \mathbf{J}_{\bm{s}}\\
    & =  (\tilde{\mathbf{B}}^T\bm{\rho}^T - \tilde{\mathbf{A}}^T)(\mathbf{I} -\bm{\rho}\bm{\rho}^T )^{-1}(\tilde{\mathbf{B}}^T\bm{\rho}^T - \tilde{\mathbf{A}}^T)^T + \tilde{\mathbf{B}}^T\tilde{\mathbf{B}} + \mathbf{J}_{\bm{s}}
\end{array}
\label{eq:InfoFunTwoVsOnePreWhitened}
\end{equation}
which is only a function of the data model parameters and the cross correlation matrix $\bm{\rho}$. Apparently when the observation noises are uncorrelated we have $\bm{\rho} = \mathbf{0}$, resulting in the corner case
\begin{equation}
{\mathbf{J}_{\bm{x},\bm{y}}}|_{\bm{\rho} = \mathbf{0}}=   \tilde{\mathbf{A}}^T\tilde{\mathbf{A}} + \tilde{\mathbf{B}}^T\tilde{\mathbf{B}} + \mathbf{J}_{\bm{s}}
\label{eq:InfoFunTwoVsOnePreWhitenedZeroCorr}
\end{equation}

Noting that $(\mathbf{I} -\bm{\rho}^T\bm{\rho} ) \succ \mathbf{0}$, the eigenvalues of $\bm{\rho}\bm{\rho}^T$ are all smaller than one\footnote{Note that interpreting $(\mathbf{I}-\bm{\rho}\bm{\rho}^T)$ as a projection matrix provides additional insights for the interpretation of the joint information of the two sensor sets. 
}. Therefore, another corner case is when $\bm{\rho}\bm{\rho}^T \rightarrow \mathbf{I}$, or its eigenvalues approach unity. In this case, if the other factors in \eqref{eq:InfoFunTwoVsOnePreWhitened} are non-zero, the Fisher information becomes infinite and the MMSE is zero. This practically means that perfect noise rejection can be obtained by merging the two sensor sets (modalities), when the noises are fully correlated.
\subsubsection{Optimal multi-sensor set cross-correlation}
Another question concerns the optimal cross correlation between the noises of the two modalities that would result in the best synergy between the two modalities. Optimizing \eqref{eq:InfoFunTwoVsOnePreWhitened} with respect to the cross correlation function can be performed by maximizing:
\begin{equation}
e \stackrel{\Delta}{=} \Tr(\mathbf{J}_{\bm{x},\bm{y}})
    \label{eq:synergycost}
\end{equation}
which is the inverse of the minimum mean square error. The maximizers of \eqref{eq:synergycost} are the roots of:
\begin{equation}
\begin{array}{rl}
\displaystyle\frac{\partial e}{\partial \bm{\rho}} &= 2[\tilde{\mathbf{A}} + \bm{\rho}(\mathbf{I} - \bm{\rho}^T\bm{\rho})^{-1}(\tilde{\mathbf{A}}^T\bm{\rho} - \tilde{\mathbf{B}}^T)^T](\tilde{\mathbf{A}}^T\bm{\rho} - \tilde{\mathbf{B}}^T)(\mathbf{I} - \bm{\rho}^T\bm{\rho})^{-1}\\
&= 2[\tilde{\mathbf{B}} + \bm{\rho}^T(\mathbf{I} - \bm{\rho}\bm{\rho}^T)^{-1}(\tilde{\mathbf{B}}^T\bm{\rho}^T - \tilde{\mathbf{A}}^T)^T](\tilde{\mathbf{B}}^T\bm{\rho}^T - \tilde{\mathbf{A}}^T)(\mathbf{I} - \bm{\rho}\bm{\rho}^T)^{-1}
\end{array}
\label{eq:synergycostgrad}
\end{equation}
Apparently, $\tilde{\mathbf{B}}=\bm{\rho}^T\tilde{\mathbf{A}}$ and $\tilde{\mathbf{A}}=\bm{\rho}\tilde{\mathbf{B}}$ are zeros of $\frac{\partial e}{\partial \bm{\rho}}$.  
In practice, since the absolute singular values of $\bm{\rho}$ are smaller than one, admissibility conditions are required for the existence of these roots. In fact, since $0 < \lambda(\bm{\rho}\bm{\rho}^T) < 1$, only the modality with a weaker SNR may be redundant (can be shown by calculating the SNR matrices $\tilde{\mathbf{A}}^T\tilde{\mathbf{A}}$ and $\tilde{\mathbf{B}}^T\tilde{\mathbf{B}}$ in the above cases). We summarize the results on the synergy between two set of sensors in Lemma \ref{theorem:twochannelsynnergy}.
\begin{center}
\noindent\fbox{%
    \parbox{\textwidth}{%
\begin{lemma}
\label{theorem:twochannelsynnergy}
For two sets of pre-whitened observations $\tilde{\bm{x}}=\tilde{\mathbf{A}}\bm{s}+\tilde{\bm{v}}$ and $\tilde{\bm{y}}=\tilde{\mathbf{B}}\bm{s}+\tilde{\bm{u}}$, where $\tilde{\bm{v}}\sim\mathcal{N}(\mathbf{0}, \mathbf{I})$, $\tilde{\bm{u}}\sim\mathcal{N}(\mathbf{0}, \mathbf{I})$ and $\mathbb{E}(\tilde{\bm{v}}\tilde{\bm{u}})=\bm{\rho}$, multiple scenarios can be considered resulting in using both or one of the modalities:
\begin{itemize}
    \item Modality selection applications: if $\tilde{\mathbf{A}}^T\tilde{\mathbf{A}}\succ\tilde{\mathbf{B}}^T\tilde{\mathbf{B}}$, using $\tilde{\bm{x}}$ alone results in a smaller MMSE estimation error in all entries of the vector $\bm{s}$, as compared to using $\tilde{\bm{y}}$; otherwise, there exist one or more entries of $\bm{s}$ for which using $\tilde{\bm{y}}$ would result in smaller or equal MMSE;
    \item Modality fusion applications: in terms of Fisher information, using both observations is always preferred over (or is at least as good as), using any single observation;
    \item Weakly correlated noise: if $\bm{\rho} = \mathbf{0}$, the Fisher information of the two modalities are additive ($\tilde{\mathbf{A}}^T\tilde{\mathbf{A}} + \tilde{\mathbf{B}}^T\tilde{\mathbf{B}}$);
    \item Strongly correlated noise: if $\bm{\rho}$ approaches a unitary matrix ($\bm{\rho}\bm{\rho}^T \rightarrow \mathbf{I}$) and $\tilde{\mathbf{A}}\neq\bm{\rho}\tilde{\mathbf{B}}$ and $\tilde{\mathbf{B}}\neq\bm{\rho}^T\tilde{\mathbf{A}}$, the Fisher information approaches infinity (the MMSE attains its minimum);
    \item Modality redundancy: if $\tilde{\mathbf{B}}=\bm{\rho}^T\tilde{\mathbf{A}}$, then $\tilde{\bm{y}}$ is redundant and the Fisher information is $\tilde{\mathbf{A}}^T\tilde{\mathbf{A}}$, and if $\tilde{\mathbf{A}}=\bm{\rho}\tilde{\mathbf{B}}$, then $\tilde{\bm{x}}$ is redundant and the Fisher information is $\tilde{\mathbf{B}}^T\tilde{\mathbf{B}}$;
    \item Admissibility: $(\mathbf{I} -\bm{\rho}^T\bm{\rho} ) \succ \mathbf{0}$ and $|\lambda(\bm{\rho}\bm{\rho}^T)|< 1$; hence, only the modality with a weaker SNR may be redundant (the smaller between $|\tilde{\mathbf{A}}^T\tilde{\mathbf{A}}|$ and $|\tilde{\mathbf{B}}^T\tilde{\mathbf{B}}|$)
\end{itemize}
\end{lemma}
}%
}
\end{center}
Some of the special two scalar sensor cases studied in \citep{chlaily2018modele} are special cases of the above lemma.

\subsection{Optimal secondary sensor configuration}
\label{sec:optimalplacement}
Another interesting scenario is when we have already a set of (pre-whitened) sensors $\tilde{\bm{x}}=\tilde{\mathbf{A}}\bm{s}+\tilde{\bm{v}}$ and would like to set up a secondary set of sensors $\tilde{\bm{y}}=\tilde{\mathbf{B}}\bm{s}+\tilde{\bm{u}}$, with maximum synergy with the first group of sensors. In other words, we seek the secondary sensor configuration with minimal redundancy with the first set. The answer to this problem is nontrivial. It has recently gained notable attention in the blind source separation literature \cite{Ghayem2019,Ghayem2021}. Most recently, a game theoretical approach based on SHAP (SHapley Additive exPlanations) values has been proposed \cite{Pelegrina2023}.

The hereby proposed information-based approach can provide an alternative perspective to this nontrivial problem. Considering that the first sensor set has been readily fixed, the problem can be formulated as a matrix design problem for the secondary mixing matrix $\tilde{\mathbf{B}}$. Assuming that the secondary set of sensors has a bounded SNR, say $\Tr(\tilde{\mathbf{B}^T}\tilde{\mathbf{B}}) \leq p$, the problem can be formulated as follows:
\begin{equation}
    \tilde{\mathbf{B}}^* =\argmax_{\tilde{\mathbf{B}}}[\Tr(\mathbf{J}_{\bm{x},\bm{y}}) - \lambda \Tr(\tilde{\mathbf{B}}^T\tilde{\mathbf{B}})]
\label{eq:SecondaryConfig}
\end{equation}
for some $\lambda\geq 0$, which by differentiation with respect to $\tilde{\mathbf{B}}$ results in
\begin{equation}
    \tilde{\mathbf{B}}^* = [\mathbf{I} - \lambda(\mathbf{I} -\bm{\rho}^T\bm{\rho})]^{-1}\bm{\rho}^T\tilde{\mathbf{A}}
\label{eq:OptSecondaryConfig}
\end{equation}
and the Lagrange multiplier $\lambda$ is selected such that the SNR constraint $\Tr(\tilde{\mathbf{B}^T}\tilde{\mathbf{B}}) \leq p$ is fulfilled, or remains as a design parameter, if $p$ is unknown\footnote{Finding $\lambda$ for a known $p$ can be achieved by performing the SVD $\bm{\rho}=\mathbf{U}\bm{\Sigma}\mathbf{V}^T$. Using this method, we can show that $\lambda$ is the root of:
\begin{equation}
    \displaystyle \sum_i \frac{d_{ii}\sigma_i^2}{[1 - \lambda(1-\sigma_i^2)]^2} = p
\end{equation}
where $\sigma_i$ are the singular values of $\bm{\rho}$ and $d_{ii}$ are the diagonal entries of $\mathbf{U}^T\tilde{\mathbf{A}}\tilde{\mathbf{A}}^T\mathbf{U}$. This equation can be solved by \textit{root locus} approaches common in linear control theory. 
}. The optimal secondary configuration, clearly depends on the primary configuration, the correlation between the noises of both configurations, and the SNR ``budget''. In the corner case that $\bm{\rho}^T\bm{\rho}=\mathbf{I}$, $\lambda$ vanishes and $\tilde{\mathbf{B}}=\bm{\rho}^T\tilde{\mathbf{A}}$ would be the optimal sensor configuration, regardless of the secondary SNR, which is equivalent to the modality redundancy case found in Lemma \ref{theorem:twochannelsynnergy}. Equation \eqref{eq:OptSecondaryConfig} can be further simplified using the singular value decomposition of $\bm{\rho}$.

\subsection{CRLB for the nonlinear data model case}
\label{sec:nonlinearcase}
Some of the properties derived in previous sections are extendable to nonlinear data models contaminated by additive noise. Let us consider the nonlinear data model
\begin{equation}
    \bm{x} = \mathbf{h}(\bm{s}) + \bm{v}
\label{eq:nonlinearmixture}
\end{equation}
where $\bm{v}\sim\mathcal{N}(\mathbf{0}, \bm{\Sigma})$ is the measurement noise, independent from $\bm{s}$. In this case, the log-likelihood function is
\begin{equation}
    ll(\bm{x}|\bm{s}) \stackrel{\Delta}{=}\log f_{\bm{X}|\bm{S}}(\bm{x}|\bm{s}) = -\frac{1}{2}\log( (2\pi)^{n}\det(\bm{\Sigma})) -\frac{1}{2}(\bm{x}-\mathbf{h}(\bm{s}))^T\bm{\Sigma}^{-1}(\bm{x}-\mathbf{h}(\bm{s}))
    \label{eq:NonLinLogLikelihood}
\end{equation}
and
\begin{equation}
    \frac{\partial}{\partial \bm{s}} ll(\bm{x}|\bm{s}) = \bm{\nabla}_{\mathbf{h}}(\bm{s})\bm{\Sigma}^{-1}(\bm{x}-\mathbf{h}(\bm{s}))
    \label{eq:NonLinLogLikelihoodPartial}
\end{equation}
where $\bm{\nabla}_{\mathbf{h}}(\bm{s})$ denotes the Jacobian matrix of the nonlinear transforms. Using the same procedure as in Section \ref{sec:MLCRLB}, results in
\begin{equation}
\mathbf{J}_{\bm{x}|\bm{s}} = \mathbb{E}_{\bm{s}}\{\bm{\nabla}_{\mathbf{h}}(\bm{s})\bm{\Sigma}^{-1}\bm{\nabla}_{\mathbf{h}}(\bm{s})^T\}
\label{eq:fisherinformationGaussiannonlinear}
\end{equation}
which is the nonlinear extension of the Fisher information derived in \eqref{eq:fisherinformationGaussian}. Similarly, we obtain the following information matrix for the case in which the source vector $\bm{s}$ is a random vector:
\begin{equation}
    \mathbf{J}_{\bm{x}} = \mathbb{E}_{\bm{s}}\{\bm{\nabla}_{\mathbf{h}}(\bm{s})\bm{\Sigma}^{-1}\bm{\nabla}_{\mathbf{h}}(\bm{s})^T\} + \mathbf{J}_{\bm{s}}
    \label{eq:totalinformationMatrixNonLin}
\end{equation}
which is the nonlinear extension of \eqref{eq:totalinformationMatrix} --- and the equivalent of the SNR matrix. This result shows that in the nonlinear case, the information obtained by the observation and priors are again additive; but this time, both depend on the source distribution. As a result, the model and prior effects are not separable in the nonlinear case.

\subsection{Fisher information for two sets of nonlinear mixtures}
\label{sec:crlbtwogroupnonlin}
We now move to the case of two nonlinear modalities
\begin{equation}
\begin{array}{cc}
\bm{x} = \mathbf{h}(\bm{s}) + \bm{v}\\
\bm{y} = \mathbf{g}(\bm{s}) + \bm{u}
\end{array}
\label{eq:nonlinearmixtureTwoGroup}
\end{equation}
where $\bm{v}\sim\mathcal{N}(\mathbf{0}, \bm{\Sigma}_v)$ and $\bm{u}\sim\mathcal{N}(\mathbf{0}, \bm{\Sigma}_u)$ are Gaussian measurement noise, independent from $\bm{s}$, with a joint covariance matrix as previously defined in \eqref{eq:noisecovariance}. As in \eqref{eq:whitened}, using the decompositions $\bm{\Sigma}_{u}=\bm{L}_u\bm{L}_u^T$ and $\bm{\Sigma}_{v}=\bm{L}_v\bm{L}_v^T$, the whitened version of the observations are
\begin{equation}
\begin{array}{l}
\tilde{\bm{x}}=\bm{L}_v^{-1}\bm{x}=\bm{L}_v^{-1}\mathbf{h}(\bm{s})+\bm{L}_v^{-1}\bm{v} = \tilde{\mathbf{h}}(\bm{s})+\tilde{\bm{v}}\\
\tilde{\bm{y}}=\bm{L}_u^{-1}\bm{y}=\bm{L}_u^{-1}\mathbf{g}(\bm{s})+\bm{L}_u^{-1}\bm{u} = \tilde{\mathbf{g}}(\bm{s})+\tilde{\bm{u}}
\end{array}
\label{eq:nonlinearwhitened}
\end{equation}
where as with the linear case $\tilde{\bm{v}}\sim\mathcal{N}(\mathbf{0}, \mathbf{I})$, $\tilde{\bm{u}}\sim\mathcal{N}(\mathbf{0}, \mathbf{I})$ and $\bm{\rho}=\mathbb{E}(\tilde{\bm{v}}\tilde{\bm{u}}^T)=\bm{L}_v^{-1}\bm{\Sigma}_{vu}\bm{L}_u^{-T}$ (or equivalently $\mathbb{E}(\tilde{\bm{u}}\tilde{\bm{v}}^T)=\bm{\rho}^T=\bm{L}_u^{-1}\bm{\Sigma}_{uv}\bm{L}_v^{-T}$). Therefore, we find
\begin{equation}
\begin{array}{rl}
    \mathbf{J}_{\bm{x}} = & \mathbb{E}_{\bm{s}}\{\bm{\nabla}_{\tilde{\mathbf{h}}}(\bm{s})\bm{\nabla}_{\tilde{\mathbf{h}}}(\bm{s})^T\} + \mathbf{J}_{\bm{s}}\\
    \mathbf{J}_{\bm{y}} = & \mathbb{E}_{\bm{s}}\{\bm{\nabla}_{\tilde{\mathbf{g}}}(\bm{s})\bm{\nabla}_{\tilde{\mathbf{g}}}(\bm{s})^T\} + \mathbf{J}_{\bm{s}}\\
    \mathbf{J}_{\bm{x},\bm{y}} = &  \mathbb{E}_{\bm{s}}\{(\bm{\nabla}_{\tilde{\mathbf{h}}}(\bm{s})\bm{\rho} - \bm{\nabla}_{\tilde{\mathbf{g}}}(\bm{s}))(\mathbf{I} -\bm{\rho}^T\bm{\rho} )^{-1}(\bm{\nabla}_{\tilde{\mathbf{h}}}(\bm{s})\bm{\rho} - \bm{\nabla}_{\tilde{\mathbf{g}}}(\bm{s}))^T\} + \mathbb{E}_{\bm{s}}\{\bm{\nabla}_{\tilde{\mathbf{h}}}(\bm{s})\bm{\nabla}_{\tilde{\mathbf{h}}}(\bm{s})^T\} + \mathbf{J}_{\bm{s}}\\
    =&  \mathbb{E}_{\bm{s}}\{(\bm{\nabla}_{\tilde{\mathbf{g}}}(\bm{s})\bm{\rho}^T - \bm{\nabla}_{\tilde{\mathbf{h}}}(\bm{s}))(\mathbf{I} -\bm{\rho}\bm{\rho}^T )^{-1}(\bm{\nabla}_{\tilde{\mathbf{g}}}(\bm{s})\bm{\rho}^T - \bm{\nabla}_{\tilde{\mathbf{h}}}(\bm{s}))^T\} + \mathbb{E}_{\bm{s}}\{\bm{\nabla}_{\tilde{\mathbf{g}}}(\bm{s})\bm{\nabla}_{\tilde{\mathbf{g}}}(\bm{s})^T\} + \mathbf{J}_{\bm{s}}
\end{array}
\label{eq:nonlinInfoFunTwoLinMix}
\end{equation}
By comparing these results with the linear case, we notice that statements similar to Lemma \ref{theorem:twochannelsynnergy} exist in the nonlinear case, this time using the Jacobian of the nonlinear mixture.

\section{Conclusion and future work}
In this work we provided theoretical insights to help understanding of sensor fusion and selection in multimodal and multichannel systems. Grounded in information and estimation theory, we investigated linear and nonlinear signal mixtures contaminated by additive (Gaussian) noise. This is a recurrent problem in many applications, including biomedical applications where a physiological phenomenon can be monitored using multiple electrodes or using multiple modalities. A key aspect of this study was the exploration of the multichannel signal-to-noise ratio (SNR) matrix, which emerges as a critical tool in discerning the conditions under which sensor selection or fusion is either beneficial or redundant. The practical implications of these findings are substantial, as they guide the decision-making process in designing multichannel/multimodal systems.


In future research, these preliminary results can be extended from various perspectives. The channel/modality redundancy conditions obtained in Lemma \ref{theorem:twochannelsynnergy} require further investigation, particularly for cases where $\bm{x}$ and $\bm{y}$ do not have equal lengths. The notion of the SNR matrix was instrumental in this work. Additional insights can be gained by studying the eigenstructure of the SNR matrix using singular value decomposition or by employing \textit{concentration ellipses} \citep[Sec. 2.4]{VanTrees2001detection} to provide further understanding for using the CRLB in modality selection and fusion.

From the applied perspective, we can study the impact of the hereby studied ideas for problems such as lead selection and fusion for noninvasive fetal \citep{Ghayem2021}, and adult \cite{reyna2021will,reyna2022issues} electrocardiography; or the problem of multimodal cardiac monitoring \citep{kazemnejad2021open}.


\section*{Acknowledgement}
R.~Sameni acknowledges support from the American Heart Association Innovative Project Award on Developing Multimodal Cardiac Biomarkers for Cardiovascular-related Health Assessment (23IPA1054351).

\bibliographystyle{abbrvnat}
\bibliography{References}
\end{document}